\documentclass{article}
\usepackage[utf8x]{inputenc} % allow utf-8 input
\usepackage{arxiv}
\usepackage{subfiles}
\usepackage{amsmath}
\usepackage{diagbox}

\usepackage{multirow}
\usepackage{float}
\usepackage{graphicx}
\graphicspath{{images/}{../images/}}
\usepackage{subcaption}
\usepackage[export]{adjustbox}
\usepackage{wrapfig}
\usepackage{hyperref}
\hypersetup{
    colorlinks=true,
    linkcolor=blue,
    filecolor=magenta,      
    urlcolor=cyan,
}

\title{Data Mining in Large Frequency Tables With Ontology, with an Application to the Vaccine Adverse Event Reporting System}
\author{
  Bangyao Zhao\\
  Department of Biostatistics\\
  University of Michigan\\
  Ann Arbor, MI 48109 \\
  \texttt{byzhao@umich.edu} \\
  \And 
  Lili Zhao\\
  Department of Biostatistics\\
  University of Michigan\\
  Ann Arbor, MI 48109 \\
  \texttt{zhaolili@umich.edu} \\
}
\date{August 2020}

\begin{document}

\maketitle

\begin{abstract}

Vaccine safety is a concerning problem of the public, and many signal detecting methods have been developed to identify relative risks between vaccines and adverse events (AEs). Those methods usually focus on individual AEs, where the randomness of data is high. The results often turn out to be inaccurate and lack of clinical meaning. The AE ontology contains information about biological similarity of AEs. Based on this, we extend the concept of relative risks (RRs) to AE group level, which allows the possibility of more accurate and meaningful estimation by utilizing data from the whole group. In this paper, we propose the method zGPS.AO (Zero Inflated Gamma Poisson Shrinker with AE ontology) based on the zero inflated negative binomial distribution. This model has two purples: a regression model estimating group level RRs, and a empirical bayes framework to evaluate AE level RRs. The regression part can handle both excess zeros and over dispersion in the data, and the empirical method borrows information from both group level and AE level to reduce data noise and stabilize the AE level result. We have demonstrate the unbiaseness and low variance features of our model with simulated data, and obtained meaningful results coherent with previous studies on the VAERS (Vaccine Adverse Event Reporting System) database. The proposed methods are implemented in the R package \href{https://github.com/umich-biostatistics/zGPS.AO}{zGPS.AO}, which can be installed from the Comprehensive R Archive Network, CRAN. The results on VAERS data are visualized using the interactive web app \href{https://bangyaozhao.shinyapps.io/vaers/?_ga=2.83997145.1917783035.1602556883-470105323.1595624541}{Rshiny}. 

\end{abstract}

% keywords can be removed

\section{Introduction}

The Centers for Disease Control and Prevention (CDC) and the U.S. Food and Drug Administration (FDA) conduct post-licensure vaccine safety monitoring using the Vaccine Adverse Event Reporting System (VAERS) \cite{Varricchio:2004,Shimabukuro:2015}. VAERS accepts spontaneous reports of suspected vaccine adverse events after administration of any vaccine licensed in the United States from 1990 to present. As a national public health surveillance resource, VAERS is a key component in ensuring the safety of vaccines. 

Numerous methods have been used to conduct safety studies with the VAERS database \cite{DuMouchel:1999,Evans:2001,van:2002,Bate:1998,Orre:2000,Noren:2006,DuMouchel:2001,Szarfman:2002,Kulldorff:2011,Davis:2005,Li:2009,Li:2014,Kulldorff:2013}. In these methods, a contingency table is generally created to display counts for all vaccine and adverse event pairs during a specified time period. In this table, each row represents a vaccine and each column represents an adverse event (AE). Each cell in the table contains the number of VAERS reports that mention both that vaccine and that event for a defined period. A statistical measure is then calculated to quantify the association between an adverse event and a vaccine. A large value of the measure shows a strong association, which might indicate a vaccine safety problem (called ``signal"). A signal is considered evidence that an adverse event might be caused by vaccination and warrants further investigation or action. However, these methods frequently identify many AE signals and they are often hard to interpret in a biological context.

\section{Method}

\subsection{Statistical Model}

%The VAERS report includes patient demographic information, vaccination date, vaccine type(s), adverse event(s), and other relevant information, and one report often involves more than one vaccine and may involve more than one reported adverse event. Here 

To mining for unusually frequent vaccine-AE combinations, we summarize  VAERS reports into a large AE-vaccine contingency table of $I$ rows (vaccines) and $J$ columns (AEs) (see Table \ref{table:full_tb}). The cell count $y_{ij}$ is the total number of reports mentioned both vaccine $i$ and the AE $j$ in a specific time period, the column margin $y_{i.}=\sum_{j=1}^{J}y_{ij}$ is the total number of reports mentioned the vaccine $i$, the row margin $y_{.j}=\sum_{i=1}^{I}y_{ij}$ is the total number of reports mentioned the AE $j,$ and $y_{..}=\sum_{ij}y_{ij}$ is the total number of reports in this time period.

\begin{table}[H]
    \centering
        \begin{tabular}{|c |c c c c|} 
            \hline
            \diagbox{Vaccine}{AE} & $AE_1$ & $AE_2$ & ... & $AE_J$ \\ 
            \hline
            $VAX_1$& $y_{11}$ & $y_{12}$ & ...& $y_{1J}$ \\ 
             \hline
             $VAX_2$&$y_{21}$ & $y_{22}$ & ...& $y_{2J}$ \\
            \hline
            ...&... & ... & ... & ... \\
            \hline
             $VAX_I$&$y_{I1}$ & $y_{I2}$ & ...& $y_{IJ}$ \\
            \hline
        \end{tabular}
    \caption{An example of the contingency table}
    \label{table:full_tb}
\end{table}

The strength of the vaccine-AE association is commonly expressed in terms
of a ratio of the observed to the expected frequencies of report. Based on \cite{DuMouchel:1999}, the expected frequency for the $ij$ vaccine-AE pair is defined as $M_{ij}=\frac{y_{i.}y_{.j}}{y_{..}},$ which is the frequency we would observe if vaccine and AE are independent. The relative reporting rate (RR) is defined as $RR_{ij}=\frac{y_{ij}}{M_{ij}}.$ If RR$=3.2$ for a
vaccine-AE pair, then this pair occurred in the data 3.2 times more frequently than expected under the assumption of no association between the vaccine and the AE. In \cite{DuMouchel:1999}, $y_{ij}$ is assumed to have a Poisson distribution, i.e.,  $y_{ij} \sim Poisson(M_{ij} \lambda_{ij}),$ where $\lambda_{ij}$ is the parameter of interest for estimating RR, with  a larger value  indicating a stronger association between vaccine $i$ and AE $j$. 

\cite{DuMouchel:1999} enhances the simple use of the separate Poisson model by
allowing "shrinkage" of similar $\lambda_{ij}$'s towards each other, thereby reducing the effect of sampling variation in the data. In this paper,  we define similar AEs using the AE ontology in MedDRA (Medical Dictionary for Regulatory Activities) [reference] \cite{Mozzicato:2009}. MedDRA describe AE relationships by a five level hierarchy.  VAERS uses the second lowest term, ``Preferred Terms" (PT), which is a distinct descriptor for a symptom, sign and disease. Related PTs are grouped into higher-level AE terms, including ``High Level Group Terms" (HLGT) and ``System Organ Classes" (SOC). By incorporating the AE ontology in modelling, we allow related AEs shrinking towards each other. Specifically, we assume that $\lambda_{ij}$ parameters in the same AE group are generated from a common gamma distribution. This common distribution allows information sharing between $\lambda$'s in the same AE group. The gamma-Poisson mixture distribution is equivalent to a negative binomial distribution for $y_{ij}$. To accommodate excessive zero frequencies for vaccine-pairs in VAERS dataset, we consider a zero-inflated negative binomial distribution (ZINB) for modelling a group of AE counts. Under this modelling framework, we are able to mining safety signals for both the AE groups and individual AEs within the each group simultaneously.

%$RR_{ij}$ is the Maximum Likelihood Estimator (MLE) for $\lambda_{ij}$.

\subsection{Mining safety signal for AE groups}

We model each AE group separately.  Suppose an AE group include $K$ AE terms, $\{AE_1,..,AE_K\},$ and we assume that $y_{ik}\sim ZINB(r,p_i,M_{ik}\mu_i)$, ($i=1,\cdots,I; k=1,\cdots,K$), 

%We assume these counts follow a zero inflated negative binomial distribution (ZINB), with a common dispersion parameter $r$, while the zero part probability $p$ and the log count mean parameter $\varph_i$ differ by vaccine types.
% TODO ADD INDICES and offset
% \begin{equation}
%     ZINB(y_{ik}|r,p_i,\varphi_i;M_{ik})=
%     \begin{cases}
%       p_i+(1-p_i)NB(0|r,\varphi_i;M_{ik}) & \text{if}\ y_{ik}=0 \\
%       (1-p_i)NB(y_{ik}|r,\varphi_i;M_{ik}) & \text{if}\ y_{ik}>0,
%     \end{cases}
% \end{equation}
\begin{equation}
    ZINB(y_{ik}|r,p_i,M_{ik}\mu_i)=
    \begin{cases}
      p_i+(1-p_i)NB(0|r,M_{ik}\mu_i) & \text{if}\ y_{ik}=0 \\
      (1-p_i)NB(y_{ik}|r,M_{ik}\mu_i) & \text{if}\ y_{ik}>0,
    \end{cases}
\end{equation}
where $p_i$ is the proportion of zeros and $NB(r,M_{ik}\mu_i)$ denotes a negative binomial (NB) with the dispersion parameter $r$ and the mean $M_{ik}\mu_i$. The dispersion parameter, $r,$ is assumed to be the same for all vaccines,  and $\mu_i$ is the mean of the NB distribution when $M_{ik}=1$, with a different value for each vaccine. Under this formulation, the RR of this AE group for vaccine $i$ is defined as 

\begin{equation}
    s_i=E(y_{ik}/M_{ik})=(1-p_i)\mu_i
    \label{eq:def of s}
\end{equation}

When $p_i$ is small (i.e., the AE group has a small percentage of zero counts) and $\mu_i$ is large (i.e., a large mean for the remaining NB distributed counts), $s_i$ is large.  A large $s_i$ indicates a a higher risk of the AE group associated with vaccine $i$.

\subsection{Estimation of RR for each individual AE}

Next, we use the Empirical Bayes approach to estimate RR of each individual AE in the vaccine group. In this approach, parameter estimates ($\hat{p}$,$\hat{r}$,$\hat{\mu}$) from the ZINB model are considered as prior information. The parameter of interest for estimating individual AE is $\lambda$. In this section, we drop the indices for vaccine to ease notation burden.

The NB distribution can be represented as a gamma-Poisson mixture distribution
$$y_{k}\sim Poi(M_{k}\lambda_{k})$$
$$\lambda_{k}\sim Gamma(r,\mu /r)$$

Under the ZINB model, we generalize the Poisson distribution of $y_{ik}$ by including $\lambda_{ik}=0,$ such that

\begin{equation}
        f(y_{k}|\lambda_{k})=\frac{exp(-M_{k}\lambda_{k})(M_{k}\lambda_{k})^{y_{k}}}{y_{k}!} \mbox{~~where~~}  \lambda_{k} \geq 0
        \label{eq:likelihood}
\end{equation}

In this distribution, $y_{ik}$ is degenerated at 0 when $\lambda_{ik}=0$, which can be easily seen based on two mathematical conventions, $0!=1$ and $0^0=1$.

%to denote the degenerated r.v. 0. As we can see that equation \ref{eq:pois_pmf} is  a valid p.m.f for the degenerated r.v. 0 based on $0!=1$ and $0^0=1$. (QUESTION R.V?)
%if we admit two commonly used mathematical conventions, $0!=1$ and $0^0=1$, the equation \ref{eq:pois_pmf} is still a valid p.m.f for the degenerated r.v. 0. Therefore, in seek of mathematical conciseness, we extend the Poisson family by adding $Pois(0)$ to denote the degenerated r.v. 0.

Under this formulation, $\lambda_{k}$ can be parameterized a mixture of the degenerated random variable at 0 and a gamma distribution, 

\begin{equation}
    \pi(\lambda_{k}|p,r,\beta)=p \delta(\lambda_{k})+(1-p)
\Gamma(\lambda_{k}|r,\mu/r)
\label{eq:prior}
\end{equation}

where $\beta=e^\varphi,$ and $p,$ $r,$ and $\varphi$ are parameters in ZINB model as described in equation (1);  $\delta(\cdot)$ is the Dirac delta function, denoting the p.m.f. of the degenerated random variable at 0 \cite{Dirac}.

The posterior distribution of $\lambda_{k}$  is also as a mixture distribution,

\begin{equation}
    \pi(\lambda_{k}|y_{k}) =
    \begin{cases}
      \hat{\pi}\delta(\lambda_{k})+(1-\hat{\pi})\Gamma(\lambda_{k}|r,\frac{\beta}{r+M_{k}\mu}) & \text{if}\ y_{k}=0 \\
      \Gamma(\lambda_{k}|r+y_{k},\frac{\mu}{r+M_{k}\mu}) & \text{if}\ y_{k}>0
    \end{cases}
    , \mbox{~~where~~} \hat{\pi}=\frac{p}{p+(1-p)(\frac{r}{r+M_{k}\mu})^r}
\end{equation}
If $y_{k}=0$, the posterior distribution of $\lambda_{ik}$ is a mixture of 0 and the gamma distribution. The mixing weight, $\hat{\pi},$ is the posterior probability that $y_{k}=0$ is from the zero component of ZINB.

%Here, $\hat{p_i}$ is the posterior probability that $y_{ik}=0$ is from the zero component, and $C$ is a constant, which means the posterior distribution of $\lambda_{ik}$ is a mixture of 0 and the Gamma.

% If $y_{ik}>0$,  $\lambda_{ik}$ has a gamma  distribution,
% $$\lambda_{ik}|y_{ik}>0 \sim \Gamma(r+y_{ik},\frac{\beta_i}{r+M_{ik}\beta_i}).$$

% If $y_{ik}=0$, $$\pi (\lambda_{ik}|y_{ik}=0)=\hat{p_i}\delta(\lambda_{ik})+C\lambda_{ik}^{r-1}exp\bigg(\frac{-\lambda_{ik}}{\frac{\beta_i}{r+M_{ik}\beta_i}}\bigg), \mbox{~~where~~} \hat{p_i}=\frac{p_i}{p_i+(1-p_i)(\frac{r}{r+M_{ik}\beta_i})^r}$$

Finally, the posterior mean of $\lambda_{k}$ is
\begin{equation}
    \hat{\lambda}_{k}=E(\lambda_{k}|y_{k})=
    \begin{cases}
      (1-\hat{\pi})\frac{\mu r}{r+M_{k}\mu} & \text{if}\ y_{k}=0 \\
      \frac{\mu (r+ y_{k})}{r+M_{k}\mu} & \text{if}\ y_{k}>0
    \end{cases}
    \label{eq:lambda hat}
\end{equation}

This posterior mean is  RR estimate of AE $k$.  With some algebra, we can see that this estimate is a weighted average of the sample mean,$\frac{y_{k}}{M_{k}},$ and the prior mean, $(1-p)\mu$,
\begin{equation}
    \hat{\lambda}_{k}=E(\lambda_{k}|y_{k})=w_1 \frac{y_{k}}{M_{k}} + w_2 (1-\pi)\mu
    , \mbox{~~with~~} w_1+w_2=1
\end{equation}

If $y_{k}=0$, $ w_1=1-\frac{1-\hat{\pi}}{1-p}\frac{r}{r+M_{k}\mu}.$ When $M_{k}$ goes to infinity, $w_1$ goes to 1. Here,  $M_{k}$ reflects the amount of information in the data. When it is large, the weight of the sample mean is large and RR uses more information from the data. The derivation for $y_{k}>0$ is complex (details can be found in Appendix). 

%As a simple example of observing a zero count for a vaccine-AE pair, a large expected frequency would provide a stronger evidence for the pair to safe.

%Similar result can be found when $y_{ik}>0$, with the assumption that $y_{ik}$ and $M_{ik}$ are relatively large compared to the scale of $r$. 

%The following two equations guarantees the $w_1$ and $w_2$ to be positive:

%When $\beta_i (1-p_i)<\frac{y_{ik}}{M_{ik}}$, we have
%\begin{equation}
%    \beta_i (1-p_i)<E(\lambda_{ik}|y_{ik})<\frac{y_{ik}}{M_{ik}}(1+\frac{p_i % r}{r(1-p_i)+y_{ik}})\approx\frac{y_{ik}}{M_{ik}}
%\end{equation}
%When $\frac{y_{ik}}{M_{ik}}<\beta_i (1-p_i)$, we have
%\begin{equation}
%    \frac{y_{ik}}{M_{ik}}<E(\lambda_{ik}|y_{ik})<\beta_i %(1-p_i)(1+\frac{r}{r+M_{ik}\beta_i}\frac{p_i}{1-p_i})\approx\beta_i (1-p_i)
%\end{equation}

%If $y_{ik}$ and $M_{ik}$ keep a constant ratio and go to infinity simultaneously, then $E(\lambda_{ik}|y_{ik})$ %will go to $\frac{y_{ik}}{M_{ik}}$, meaning the Bayes estimator for RR is taking information mostly from the data.

\subsubsection{Estimate RR for an AE group and each AE within the group}

% TODO the notation is not cleaer
\textbf{Estimate parameters.} To estimate the parameters $\boldsymbol{\theta}=\big(p_1,...,p_I,\beta_1,...,\beta_I,r\big)$, we apply a generalized linear regression (GLM) method, using the canonical log link function. This is done independently for each AE group. We use $\boldsymbol{Y}$ to denote observed counts in a specific AE group (of size $K$) associated with any vaccine. ($\boldsymbol{Y}$ is a column vector of length $IK$) The vector of expected mean $\boldsymbol{\mu}$ of the count part for the ZINB distribution can be expressed as:
$$ln\ \boldsymbol{\mu}=\boldsymbol{X\varphi}+ln\ \boldsymbol{M}$$ 
where $\boldsymbol{X}$ is a binary design matrix indicating which vaccine each count is associated with. ($\boldsymbol{X}$ is a $IK$ cross $I$ matrix, without the intercept column) Here $M$ is the vector of baseline frequencies with elements corresponding to elements in $\boldsymbol{Y}$ and $\boldsymbol{\mu}$. Under the log link, $\boldsymbol{\varphi}=(ln\ \beta_1,...,ln\ \beta_I)^T$.

The vector of zero part probabilities $\boldsymbol{p}$ is transformed with the logit function, and can be expressed as:
$$logit\ \boldsymbol{p} = \boldsymbol{X\alpha}$$
where the $\boldsymbol{\alpha}$ is a transformed version of $(p_1,...,p_I)$ with $\boldsymbol{\alpha}=(logit\ p_1,...,logit\ p_I)^T$

The ZINB regression method can be implemented using the function \textit{zeroinfl} from the \textit{pscl} package in R \cite{pscl}. The function will give us $(\boldsymbol{\hat{\varphi}},\boldsymbol{\hat{\alpha}},\hat{r})$ by the default Broyden–Fletcher–Goldfarb–Shanno algorithm, which is in essence an maximum likelihood estimation (MLE). Due to the functional invariance property of MLE, $\boldsymbol{\hat{\theta}}$ can be obtained by $(\hat{\beta_1},...,\hat{\beta_I})^T=exp(\boldsymbol{\hat{\varphi}})$ and $(\hat{p_1},...,\hat{p_I})^T=exp(\boldsymbol{\hat{\alpha}})/(1+exp( \boldsymbol{\hat{\alpha}}))$. With $\boldsymbol{\hat{\theta}}$ in hand, we are able to obtain both group level and AE level RRs using equation \ref{eq:def of s} and \ref{eq:lambda hat} respectively.

% We estimate the parameters $\boldsymbol{\theta}=\big(p_1,...,p_I,\varphi_1,...,\varphi_I,r\big)$ by MLE. The likelihood function can be written as the following.

%$$\boldsymbol{L(\theta)}=\prod_{i,k}L_{ik}(\boldsymbol{\theta})=\prod_{i,k}ZINB(y_{ik}|r,p_i,\varphi_i)$$
%and
%$$\boldsymbol{\hat{\theta}}=\argmax_{\theta}\boldsymbol{L(\theta)}$$

% Let $\boldsymbol{y}=(y_{11},y_{12},...,y_{1K},......,y_{I1},y_{I2},...,y_{IK}{)}^T$ be the column vector of length $IK$, $\boldsymbol{\mu}$ be the mean of the NB part of $\boldsymbol{y}$, and $\boldsymbol{M}$ be the corresponding offset. Let $\boldsymbol{X}$ be the be the $IK$ by $K$ design matrix without the intercept column indicating the type of vaccine, and $\boldsymbol{\varphi}=(\varphi_i{)}_i$ be the column vector of length $K$. Under the log link, we can express our model in a ZINB regression format.
% $$\boldsymbol{log(\mu)=X\varphi+log(M)}$$

%The proposed method can be solved by the \textit{pscl} package in R \cite{pscl}. This package can solve for the maximum likelihood estimator (MLE) using many methods. Here we choose the default one, the Broyden–Fletcher–Goldfarb–Shanno algorithm (BFGS).

% We implement the \textit{pscl} package in R \cite{pscl} to estimate parameters in our ZINB model with the default Broyden–Fletcher–Goldfarb–Shanno algorithm.

% DERIVE PARAMETER ESTIMATES, CALCULATE $s_I$ AND $\lambda$ (ADD DETAILS........)

\textbf{Estimate significance.} As we test associations for many pairs of vaccine and AE groups, we need to control for multiple comparisons. Adjustment methods, such as \cite{Thissen:2002}, can be used to control for the false discovery rate based on the p-values generated from these association tests.  In this paper, we use the method in \cite{Huang:2011,Ding:2020} and use the $maxS=\max_{i, l}s_{il}$ ($i=1,\cdots,I; l=1,\cdots, L$) as the test statistics (the maximum is taking over all $I$ vaccines and all $L$ AE groups). By using this maximum statistics, the method is conservative in detecting multiple signals and controls the overall Type I error at a given level \cite{Huang:2011,Ding:2020}.

The distribution of $maxS$ under the null ($H_0:$ no association between vaccines and AE groups) is not analytically tractable and is obtained using permutation test. However, the permutation test needs to account for the correlation between AEs mentioned on the same report. For this reason, we consider AEs observed on the same report as a single set and reshuffle the sets of AEs across reports in the permutation test. As a simple example with two reports, if one report mentioned one vaccine denoted by $V_1$ and three AEs, $a,b,c,$ and the other report mentioned two vaccines, $V_2$ and $V_3$ and two AEs, $e$ and $f$. By reshuffling the AE sets, AE set ($a,b,c$) is linked to vaccine ($V_1,V_2$ ) and AE set ($e,f$) is linked to vaccine $V_1$. By permuting sets of AEs, the correlation between AEs are reserved. For each permuted dataset, we compute a value of $maxS$. By generating $N$ permuted datasets ($N$ is generally large, say 5,000), we obtain an empirical null distribution of $N+1$ $maxS$ values (including the $maxS$ value from the observed dataset). Let $R$ denotes the rank of the observed $maxS$ in the null distribution, then the p-value is $1-R/(N+1)$. A small p-value indicates  a strong association between the vaccine and AE group. Using the null distribution of $maxS$, we can obtain the p-values of the second largest, third largest $s$, etc.

Similarly, we can obtain p-values of RR for all individual AEs, by using the maximum of $\lambda$ (here, the maximum is taking over all $I$ vaccines and all $J$ AEs) as the test statistics.

\section{Simulation}

We conducted simulation studies for a specific AE group to investigate our method with regarding to the accuracy of estimating RR for each individual AE and the AE group. We generated 20, 50, 100, or 200 AE counts (group size) from ZINB for 3 vaccines. To make simulation studies more realistic, parameters for the three vaccines ($p_i$, $\beta_i$, and $r$) are sampled from the observed VAERS data. In order to simulate the four different numbers of AE count (group size), we sampled four sets of expected counts $M_{ik}$ from the real data. RRs for the AE group are defined as in \ref{eq:def of s}. In each simulation, we generated the RRs for all individual AEs ($\lambda_{ik}$'s) based on \ref{eq:prior}, and cell counts ($y_{ik}$'s) based on \ref{eq:likelihood}.

The $\lambda_{ik}$'s and cell counts are generated for 1000 times, and $\hat{s_i}$'s are recorded after each simulation. To evaluate the model performance, we consider two commonly metrics, the observed biases (difference of observed $\hat{s_i}$'s and the true value) and the MSE (average value of observed biases). Observed biases near 0 and smaller MSEs imply better unbiasedness and accuracy of the model. 

In the simulation studies, we also increased the expected counts ($M_{ik}$'s) by multiplying all $M_{ik}$'s by 5 to see if larger $M_{ik}$'s imply better accuracy in our model. Addtionally, we generated data from ZIP (no over-dispersion) by replacing $r$ by $+\infty$, and see if our ZINB can predict well with data having no over-dispersion. 

\begin{figure}[H]
    \centering
    \includegraphics[width=1\textwidth]{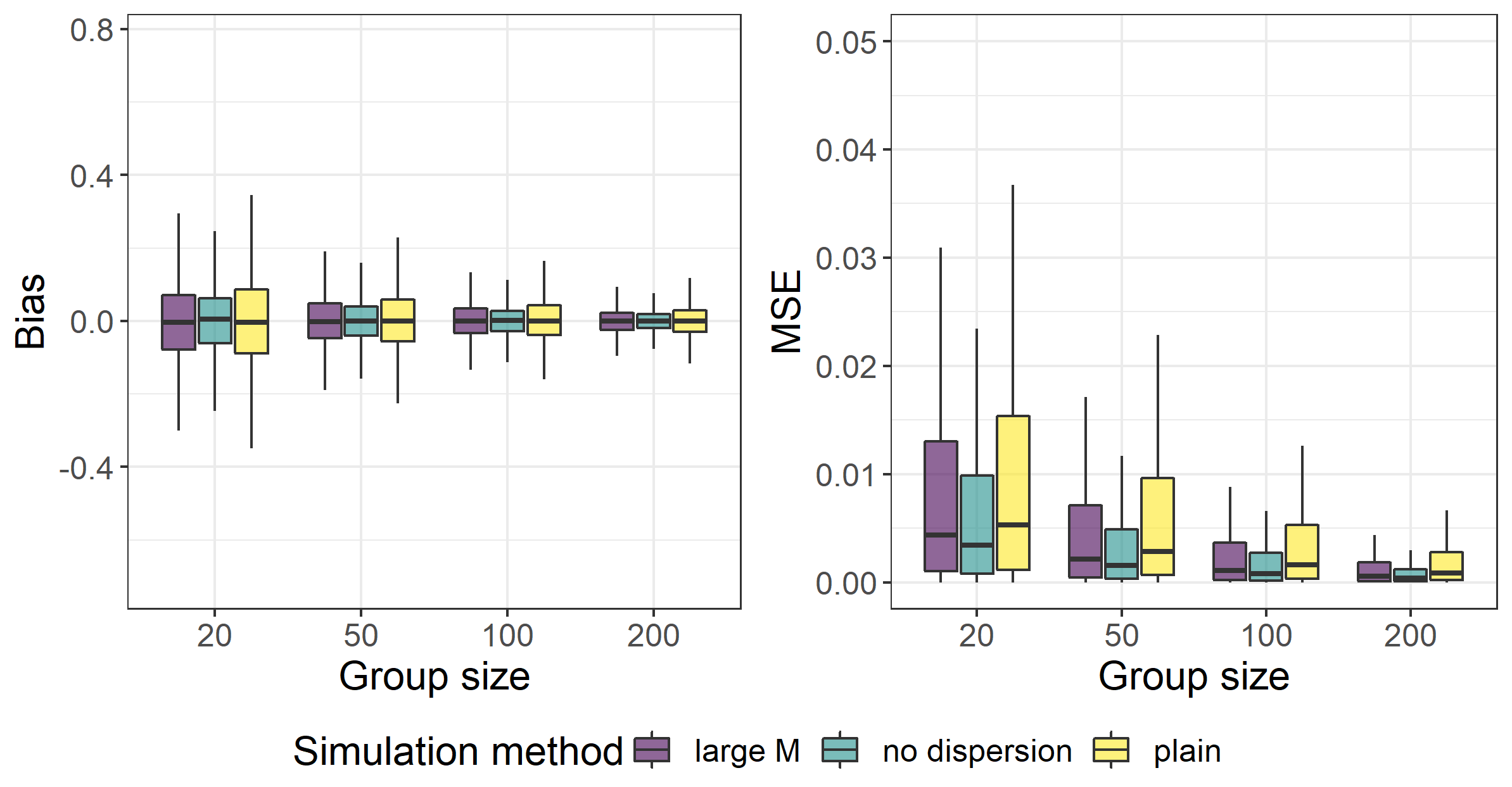}
    \caption{Group level Simulation Result: The biases are calculated as $\hat{s}_i-s_i$ for each permutation and each simulated vaccine type, and the MSE is the the squared bias.}
    \label{fig:simulation_s}
\end{figure}

Figure \ref{fig:simulation_s} shows that our method accurately estimate the RR for the AE group. an AE group of a larger size or increased expected counts had more accurate estimation of the $s$. Simulation results also demonstrate that our method can deal with data from ZIP. ZINB reduces to ZIP for data without dispersion. The same metrics on $\hat{\lambda_{ik}}$ can be used to evaluate the performance of zGPS.AO on AE level.

\begin{figure}[H]
    \centering
    \includegraphics[width=1\textwidth]{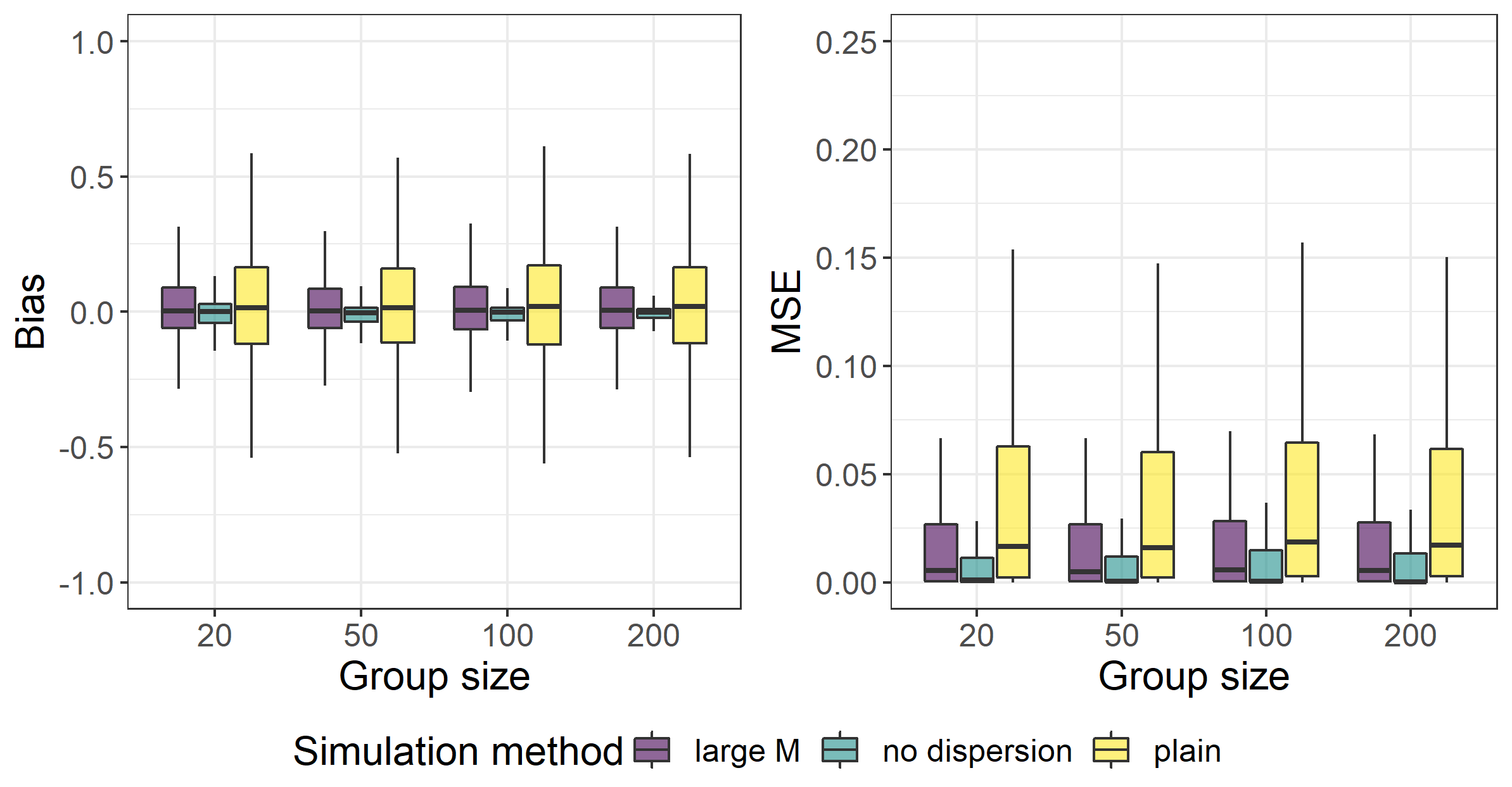}
    \caption{AE level Simulation Result: Same as on the group level, the AE level biases are calculated as $\hat{\lambda}_{ik}-\lambda_{ik}$ for each simulation, vaccine type and AE. The MSE is the squared bias.}
    \label{fig:simulation_lambda}
\end{figure}

Figure \ref{fig:simulation_lambda} shows that the AE level RR estimation in zGPS.AO is also unbiased and accurate. On the AE level, zGPS.AO can also handel data with no dispersion well, and a larger group size increases the accuracy of $\hat{\lambda_{ik}}$. The AE level estimation has higher MSE than on the group level, reflecting that fact that we have less data on AE level than on group level. 

Another finding is that, unlike what happens on group level, as the group size increases, the MSE of $\hat{\lambda}_{ik}$ stays almost at the same level. Our best explanation to this interesting phenomena is as follow: The group level estimation is quite accurate even with a moderate group size, say 20, but the cell count of a particular AE-vaccine combination is always of large randomness. Therefore, since the $\hat{\lambda}_{ik}$ borrows information from both sides, the variance of $\hat{\lambda}_{ik}$ is mainly from the randomness in the particular cell count. An increasing group size only contributes to the group level accuracy, so it does not improve the accuracy of individual $\hat{\lambda}_{ik}$ significantly.

% In each simulation, we generated 20, 50, 100, or 200 AE counts (group size) from ZINB for 3 vaccines. (WRITE ZINB MODEL). To make simulation studies more realistic, model parameters ($M$,$p_i$, $\varphi_i$, and $r$) were made similar to the real dataset (HOW? DETAILS). RR for the AE group is defined as $s=$. Then we generated the RRs for all individual AEs ($\lambda$'s) in the AE group (WRITE POISSON MODEL).

% THE FOLLOWING IS NOT CLEAR
% The $\lambda$'s and cell counts are generated for 1000 times, and $s_i$'s are recorded after each simulation. The offsets are only generated once in the beginning.

% PUT ALL THREE VACCINES INTO ONE PLOT. DON'T FIX FIGURE POSITION

% In the simulation studies, we also increased the expected counts ($M$) by multiplying all $M$ by 5. Addtionally, we generated data from ZIP (no over-dispersion) and see if our ZINB can 

% \begin{itemize}
%     \item \textbf{no\_dispersion}: Simulate data with no dispersion, i.e. $r\rightarrow +\infty$, which means %the cell counts follows a zero inflated Poisson (ZIP) distribution. 
%     \item \textbf{big\_offset}: Simulate big offsets. (We mimic this situation by multiply the sampled %offsets by 5)
% \end{itemize}

\section{Analysis of VAERS dataset}

\textbf{Process VAERS dataset.} We used reports received from 2005 to 2018 and restricted the age of the vaccine recipients between 2 to 49. We investigated AEs for 10 types of vaccine of interest, including FLU (inactivated influenza vaccine; trivalent or quadrivalent), FLUN (live attenuated influenza; trivalent or quadrivalent), HEP(Hepatitis B vaccines), HEPA (Hepatitis A vaccines), HEPAB (Hepatitis A + Hepatitis B), HPV4 (human papillomavirus 4-valent vaccine), HPV9 (human papillomavirus 9-valent vaccine), MMR (measles, mumps and rubella virus vaccine, live), TDAP (tetanus toxoid, reduced diphtheria toxoid and acellular pertussis vaccine, adsorbed), and VARCEL (Varivax-Varicella Virus, live). Those vaccines were selected from 84 types of vaccine  based on their high level of public attention and high report frequency in the dataset.  We used the HLGT level of MedDRA to define AE groups. We filtered out spurious AEs by removing AEs with a frequency less than 20 and removing AE groups containing less than 15 AEs. The final dataset for analysis has a total of 169,538 reports and 1477 AEs, which are classified to 42 AE groups.

%For a full description of those vaccine names, one may refer to the  \href{https://vaers.hhs.gov/docs/VAERSDataUseGuide_October2017.pdf}{VAERS Data Use Guide}.
A traditional way to generate vaccine-AE pair data is to get all vaccine-AE combinations in each report, regardless of the number of vaccines on the report.  If a report mentions both vaccine A and B, and an AE of fever, the current strategy creates two pairs of data: A-fever and B-fever. However, with the additional presence of vaccine B, the link might not exist between vaccine A and fever, and likewise for the link between vaccine B and fever.
We used a simple strategy proposed in \cite{Zhao:2020} to mitigate this problem. Specifically, we assign a weight to each vaccine-AE pair. If there is a single vaccine on the report, the weight is one. If there are multiple vaccines, we weight each vaccine-AE pair by the inverse of the number of vaccines mentioned in the report (i.e, $weight=\frac{1}{number of vaccines}$), assuming that the AE is linked to each vaccine with equal probability.
%traditional way is to generate all pairs of vaccine-AE combinations in each report. For example, a report with two vaccines (a and b) and one AE will generate two pairs, a-AE and b-AE. 

Under this method, the expected value of the vaccine-AE link is used as the weight. If a vaccine coexists with other vaccines in one report, the weight will be less than one, reflecting the uncertainty of the existence of the link, otherwise the weight will be one. 

\textbf{Results} 

\begin{figure}[H]
    \centering
    \includegraphics[width=1\textwidth]{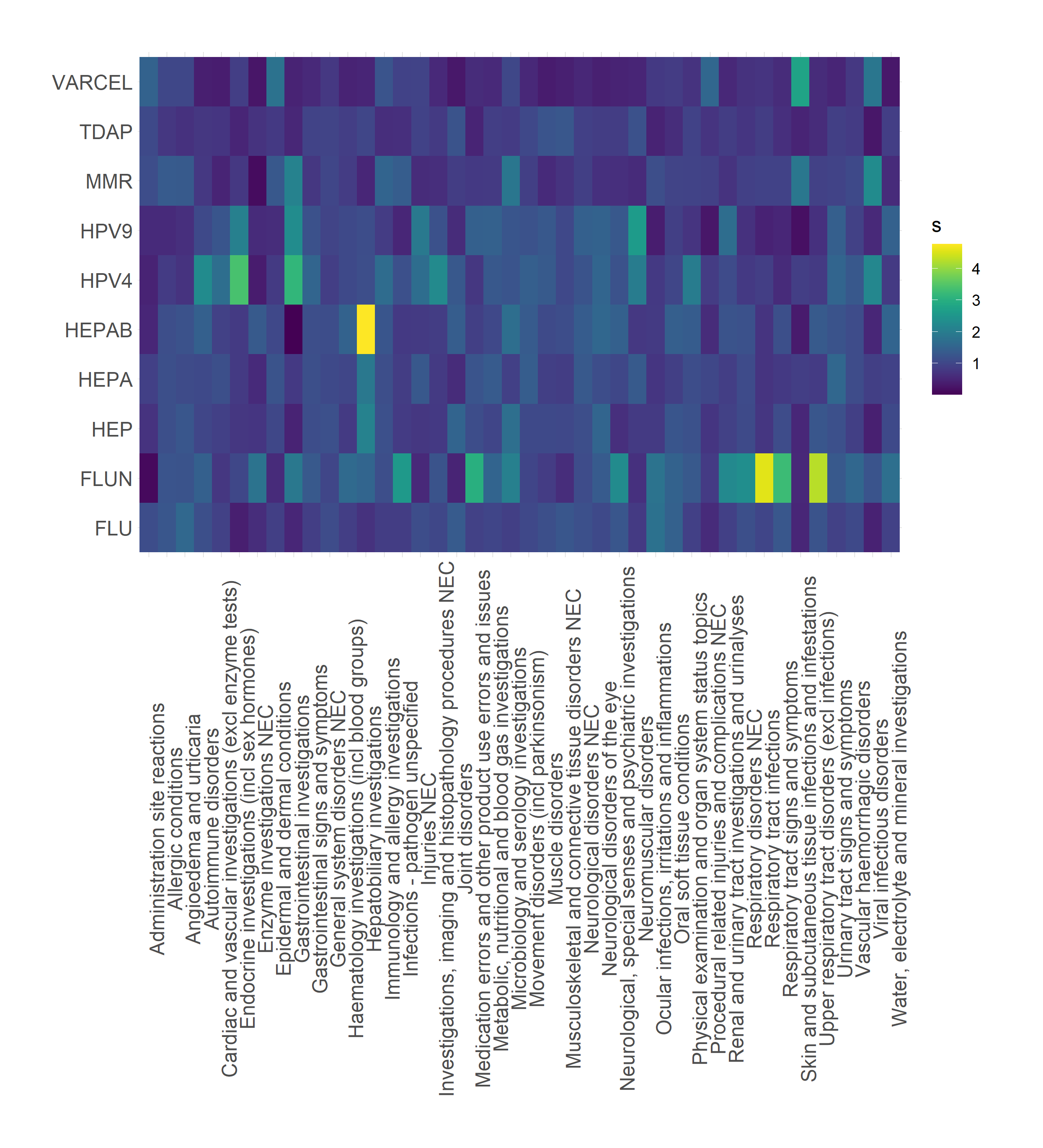}
    \caption{The heatmap of group level RRs ($s_{ij}$'s) detected in the VAERS data, with rows representing vaccine types, columns representing AE groups. The more vibrant the color is, the more intense the association is between the vaccine and the AE group.}
    \label{fig:heatmep}
\end{figure}

We applied zGPS.AO to the final dataset to investigate AEs for the 10 types of vaccine of interest.
Figure \ref{fig:heatmep} shows the result of mining AE groups by zGPS.AO. By the heatmap, we found signals on two vaccines, FLUN and HEPAB. Compared to FLU, FLUN is more  associated with respiratory related AE groups, and HEPAB is more risky compared to HEPA and HEP with the highlighted AE group Hepatobiliary investigations.

%on average, there are 1.59 vaccines and 3.82 adverse events per report.

%\paragraph{Flu vaccines} Influenza virus vaccine is designed to protect against influenza viruses that are most likely to spread and cause illness among people during the flu season. In VAERS data, two main types of influenza virus vaccines are the inactivated flu shot (FLU3 and FLU4) and the live attenuated nasal spray (FLUN3 and FLUN4). Those vaccines are widely used among population with 116349 reports containing FLU and 17093 reports containing FLUN. 
Influenza virus vaccine is designed to protect against influenza viruses that are most likely to spread and cause illness among people during the flu season. There are 116349 reports containing FLU and 17093 reports containing FLUN. We compared AE profiles of FLUN and FLU by defining a signaled AE group to have $p-value>0.01$ and $s>3$. With this definition, Table \ref{table:FLUN_significant_AE_grps} shows the 3 significant AE groups (all related to the respiratory system) associated with FLUN, while there are no significant AE groups associated with FLU. Relative to FLU, FLUN is associated with increased risk of the respiratory system. The signaled AE groups, Respiratory tract infections and Upper respiratory tract disorders have been reported before \cite{Shuoran:2020}, and the individual AEs Rhinitis, Nasal congestion, Sinus disorder are also consistent with previous researches \cite{Baxtera:2017,Haber:2015,Rob:2016}. The newly detected AE group Respiratory tract signs and symptoms is interesting and deserves more clinical trial and validation in larger dataset. 

\begin{table}[H]
\centering
\begin{tabular}{ llll }
\hline
AE group&s (p value) & top 5 AE & $\lambda$ \\ 
\hline
\multirow{5}{*}{Respiratory tract infections}&
\multirow{5}{*}{4.56 (0.001)}&
Croup infectious&9.52\\
& &Influenza&8.95\\
& &Rhinitis&6.41\\
& &Pneumonia&6.33\\
& &Atypical pneumonia&6.10\\
\hline
\multirow{5}{*}{Upper respiratory tract disorders (excl infections)}&
\multirow{5}{*}{4.26 (0.002)}&
Epistaxis&8.44\\
& &Nasal congestion&7.10\\
& &Nasal oedema&6.19\\
& &Paranasal sinus&5.53\\
& &Sinus disorder&5.22\\
\hline
\multirow{5}{*}{Respiratory tract signs and symptoms}&
\multirow{5}{*}{3.26 (0.008)}&
Nasal discomfort&9.24\\
& &Rhinorrhoea&7.29\\
& &Sneezing&6.48\\
& &Sinus headache&5.65\\
& &Rhinalgia&5.01\\
\hline
\end{tabular}
\caption{Significant AE groups associated with FLUN, while there are no AE groups associated with FLU. The top 5 significant AEs within each group is also listed here for comparison with previous results.}
\label{table:FLUN_significant_AE_grps}
\end{table}

Another finding of this study is in identifying safety problems that are likely due to interactions of two vaccines when they are administered to an individual at the same time. Specifically, we compared AE profiles induced by the hepatitis A and B combination vaccine (“Twinrix" in VAERS) to monovalent hepatitis A and B vaccines (“Havrix" for hepatitisA and “Engerix-B" for hepatitis B in VAERS). Hepatitis Vaccines have appeared 22338, 16555, and 3545 times in the dataset with Havrix, Engerix-B, and Twinrix, respectively, and are widely used in young children and teenagers to prevent liver inflammation. With the same definition of a signaled group as in studying flu vaccines, zGPS.AO detects one signaled AE group, Hepatobiliary investigations (Shown in Table \ref{table:Hepatobiliary investigations - HEP AB}), highly associated with HEPAB, while there is no signaled AE group associated with HEPA and HEP. This finding implies an increasing side effect when Hepatitis A and B vaccines are combined, and needs more statistical and clinical justification.

\begin{table}[H]
\centering
\begin{tabular}{ llll }
\hline
AE group&s (p value) & top 5 AE & $\lambda$ \\ 
\hline
\multirow{5}{*}{Hepatobiliary investigations}&
\multirow{5}{*}{4.76 (0.001)}&
Aspartate aminotransferase increased&4.81\\
& &Hepatic enzyme increased&4.81\\
& &Liver function test abnormal&4.79\\
& &Alanine aminotransferase increased&4.78\\
& &Bilirubin urine&4.78\\
\hline
\end{tabular}
\caption{The AE group hepatobiliary investigations is the only signaled group associated with HEP AB, while there are no AE groups associated with either HEP or HEP A.}
\label{table:Hepatobiliary investigations - HEP AB}
\end{table}

\section{Discussion}

In this article, we develop a method for detecting AE-vaccine associations. We assume a seperate Poisson model for each individual AE, and a ZINB model for a group of AEs. In this way, we incorporate the concept of RR, and extend it to the AE group level. The ZINB model allows our model to handle data with excess zero counts and over-dispersion. The Empirical Bayes framework has the special feature of borrowing information from both the group and the individual AE, which reduces the AE level noise and enhances the model performance. We have demonstrated the unbiasedness and low variance of our model on both AE and group levels. By applying zGPS.AO to VAERS data, we found that FLUN is associated higher risks compared to FLU, especially on AE groups related to the respiratory system, and the combination of hepatitis A and B vaccines may have higher side effects than monovalent hepatitis A or B vaccines.

% We extend the idea of RR to group level by assuming 

% BANGYAO: ADD HIGH LEVEL SUMMARY
% A PARAGRAPH TO EMPHASIZE GOOD FEATURES OF ZINB (Lili)
% 1) Assuming an NB for a group of AEs means a separate Poisson model for each individual AE
% 1) handle zero counts
% 1) ZINB allow overdispersion, if has no dispersion, the model reduces to Zero-inflated Poisson model
% 2) Under the ZINB model, each count for a vaccine-AE pair follows  a Poisson distribution, which allows estimation of association for each vaccine-AE pair.

\bibliographystyle{unsrt}
\bibliography{main}

\end{document}